\documentclass[pre,twocolumn,showpacs,superscriptaddress]{revtex4}
\usepackage[latin1]{inputenc} 
\usepackage{graphicx}  

\begin{document} 
 
\title{A Stochastic Model for the Species Abundance Problem \\ in an
  Ecological Community} 
\date{\today}  
\author{Simone Pigolotti} 
\affiliation{International School for Advanced Studies (SISSA - ISAS),
  Trieste, Italy}
\affiliation{Istituto Nazionale per la Fisica della Materia (INFM), UR - Trieste}
\author{Alessandro Flammini}
\affiliation{Dipartimento di Fisica, Universita' di Padova, v.Marzolo 8,
Padova, Italy}
\affiliation{Istituto Nazionale per la Fisica della Materia (INFM), UR - Trieste}
\author{Amos Maritan}
\affiliation{Dipartimento di Fisica, Universita' di Padova, v.Marzolo 8,
Padova, Italy}
\affiliation{Istituto Nazionale per la Fisica della Materia (INFM), UR - Padova}

\begin{abstract}

We propose a model based on coupled multiplicative stochastic processes to understand
the dynamics of competing species in an ecosystem. This process can be
conveniently described by a Fokker-Planck equation. We provide an
analytical expression for the marginalized stationary distribution. Our
solution is found in excellent agreement with numerical simulations and
compares rather well with observational data from tropical forests.

\end{abstract}

\pacs{87.23.-n, 87.23.Cc}

\maketitle 

\section{Introduction}              
One of the the most widespread quantities employed in Ecology to describe
the biodiversity in a given ecosystem is the distribution of species abundance.
In operational terms it can be defined as the histogram of the number of
species (in a well defined temporal and geographical context) consisting of
a generic number of individuals, or, from a more theoretical perspective, as 
the probability that a generic species is composed by a certain number of
individuals. Data collected in different locations suggests that the relative
species abundance distributions show a certain degree of similarity \cite{macarthur}.
To elucidate the causes that determine the shapes of these distributions and therefore 
their similarity is a problem of the uttermost importance and not only 
of theoretical nature: to understand the motives that influences  
the relative rarity or commonness of different species can be of great
help in determining policies for the conservation of the endangered ones.

The first studies on this subject can be dated back to the '40 and are due to
Fisher \cite{fisher} and Preston \cite{preston}. Their works were focused on 
finding distributions that could fit well particular data set in an empirical way. 
In particular, Preston \cite{preston}
argued that the probability of finding species with a certain
number of of individuals $x$ should be lognormal distributed, while
Fisher \cite{fisher} proposed a function of the form $e^{-ax}/x$,
with $a<<1$, the so-called Fisher log series.

Later, MacArthur \cite{macarthur} firstly pointed out that similar distributions
are found in very different ecosystems, suggesting that the shape
of such distributions is to a large extent determined by very basic, general 
and ecosystem-independent mechanisms. This in turn hinted to the
possibility to predict the shape of such distributions with simple and
general models, without taking into account too many specific details of the
ecosystem under consideration. Several models have been proposed 
that spoused this view \cite{hubbel,sole,amos}.           
Many of them restrict to modeling a single ecological community, 
a collection of similar species that feed on the same
pool of resources in a local area. This definition implies that species belonging 
to the same community interact mainly in a competitive way: in
particular, there are no prey-predator relationships among them.
The particular case of single ecological community can be 
framed in the wider context of a neutrality hypothesis.
The concept of neutrality was
firstly introduced in the framework of a biomolecular evolution theory by Kimura
\cite{kimura}, and then extended to other fields of biology. 
In the words of Hubbell \cite{hubbel}, an ecological theory can be considered neutral
when ``{\it ...treats organisms in the community as essentially identical
in their per capita probabilities of giving birth, dying, migrating
and speciating. This neutrality is defined at the individual level, 
not the species level ...}'' 

The question whether there exist ecological communities satisfying this
assumption is still rather controversial \cite{Tilman}, therefore it is 
crucial to understand what are the consequences of this zero order hypothesis \cite{Harte}.
In the context of a neutral hypothesis it is reasonable to describe
the number of offspring to which any given individual gives place to 
as a stochastic variable. As a consequence, the number of individuals
in a species at a given time can be regarded as a  multiplicative random process. 

Here we present a model aimed at reproducing the features of 
species abundance distributions under a minimal set of assumptions:
neutrality and the possibility to describe the birth process as a 
multiplicative random processes. The model translates in a Fokker-Plank 
equation for the species abundance distribution and is amenable to  
an analytical treatment. The solutions found are compared with the experimental
data we avail of. The shape of these solutions depend on one parameter, and give in
the two limiting cases both a lognormal-like curve and the Fisher log series. The
paper is organized as follows. In the second section we will present the
model and comment the assumptions made. In the third one we will take
the continuum time limit of our model, and will provide an analytical
solution for the marginalized stationary probability distribution
function. In the last two sections we compare our results
with the experimental data and comment them.

\section{The Model}

Let us consider an ecological community consisting of a fixed number, 
$s$, of species. According to MacArthur and
Wilson theory of island biogeography \cite{mcarthur}, the number of
species in a community approaches a dynamical equilibrium
between immigration, speciation and extinction. We assume that we can
neglect the fluctuations around this equilibrium value: in our model, 
when a species go extinct, it is immediately replaced by another one. 
We also assume that the net effect of the competitive interaction between
species in the community is just to keep also the total number of
individuals in the community fixed: the resources available are enough to
support just $N$ individuals across all the species. This last assumption implies
that the populations of the species undergo a zero sum dynamics. This
hypothesis is well confirmed by experimental data
\cite{preston,mcarthur}; at the end of section III we will show that relaxing
these constraints does lead to similar conclusions in the large $N$ limit. 
We introduce the $s$ variables $x_i^t$, representing
the population of the $i$-th specie at (discrete) time $t$, with the condition:
$$
\sum_{i=1}^{s}x_i^t =N \qquad \forall t
$$
Let $P(\lambda)$ be the probability that 
an individual in the community has $\lambda$ offspring 
during one time step. Here neutrality plays a key role: our assumption
implies that $P(\lambda)$ is the same for all individuals.
The population of the $i$-th species evolves according to the
following equation:

\begin{equation}\label{pois}
x_i^{t+1}=N \frac{\sum_{k=1}^{[ x_i^t]}\lambda_{k,i}^t+b}{\sum_{j=1}^s
  \left( \sum_{k=1}^{[ x_i^t]} \lambda_{k,i}^t+b\right)  }
\end{equation}

\noindent where $[\ ]$ means the integer part. We are assuming that the existence 
of species with a non integer number of individuals is not
too drastic. This might lead to round-off problems only for rare species.  
At each time step (generation) we just
sum the number of offspring of every individual belonging to that species,
and then add a small quantity $b$. This quantity becomes relevant only
for small $x_i$, and this describes the behavior of species
near their extinction threshold. We are assuming that the net effect of
extinctions, immigration and speciation can be modeled in a simple way 
with this term, whose effect is to force the $x_i$'s to be greater than
zero. Indeed, for $b = 0$, our system admits an
absorbing state with only one $x_i$ equal to $N$ and the others equal to
$0$, the so-called monodominance \cite{hubbel}.  
Notice that species are only coupled through the denominator, that
simply preserves the normalization condition.

The number of individual of each species will be typically large, so we
apply the central limit theorem to the sum of random variables in this
equation, obtaining the following model:

\begin{equation}
x_i^{t+1}=N
\frac{\bar{\lambda}x_i^t+\sigma\sqrt{x_i^t}\xi_i^t+b}{\sum_{j=1}^s
\left(\bar{\lambda}x_j^t+\sigma\sqrt{x_j^t}\xi_j^t+b \right)}
\end{equation}

\noindent where $\bar{\lambda}$ and $\sigma$ are the mean value and the
r.m.s.d. of the distribution $P(\lambda)$, and the $\xi$'s are
uncorrelated gaussian variables with zero mean and unit variance. 

It is worth noting the relation between our model and 
the multiplicative process introduced by Kesten in
\cite{kesten}. Kesten studied
random multiplicative processes of the form $X_{t+1}=\lambda_t X_t
+b_t$, where $X_t$ is the variable and both $\lambda$ and $b$ are random
variables. He found that, depending on the mean value of $\lambda$ and
on the boundary conditions, one
retrieves a lognormal or a power-law regime. Models for ecology and
economics based on this kind of processes were proposed by Sornette
\cite{sorn} and Solomon \cite{solomon}.
In our model the number of individuals of different species can be thought as following
coupled Kesten-like processes. The coupling is a consequence of the
constrain that keeps fixed to $N$ the number of individuals in the community
and that is enforced in equation (1) by the factor $N$ and by the denominator.

\section{The Continuum Limit}

In order to obtain some analytical result, we do the continuous time
limit of this model, by introducing the time interval $dt$ in the
following way:
\begin{eqnarray}
\lambda & \rightarrow &1 + \lambda dt  \nonumber \\   
b &\rightarrow & b\ dt \\  \nonumber
\sigma &\rightarrow & \sigma\ dt  \nonumber
\end{eqnarray}

By means of this substitution, our model becomes:

\begin{equation}
x_i^{t+dt}=\frac{x_i^t+dt (\bar{\lambda}x_i^t+\sigma\sqrt{x_i^t}\xi_i^t+b)}{1+\frac{dt}{N}\sum_{j=1}^s(\bar{\lambda}x_j^t+\sigma\sqrt{x_j^t}\xi_j^t+b)}
\end{equation}

Expanding the denominator and using the fact that $\sum_j x_j=N$, we
get the Langevin equation:

\begin{equation}
\dot{x_i}= f_i(x)+\sum_{j=1}^s B_{ij}(x)\xi_j
\end{equation}

\noindent where:

\begin{eqnarray}
f_i(x_i) = b(1-\frac{s}{N}x_i) \nonumber \\
B_{ij}(\underline{x}) = (\delta_{ij}-\frac{x_i}{N})\sqrt{x_j} 
\end{eqnarray}

The Fokker-Planck equation \cite{gardiner} associated to this Langevin equation is :

\begin{equation}\label{fp}
\dot{P}(\underline{x},t)=-\sum_{i=1}^s \partial_i \left[-f_i
  P(\underline{x},t) + D \sum_j \partial_j  (g_{ji}(\underline{x})P(\underline{x},t))\right]
\end{equation}

\noindent with $D=\frac{\sigma^2}{2}$ and:

\begin{equation}
g_{ij}(\underline{x})=g_{ji}(\underline{x})=\sum_k B_{ik}B_{jk}=(\delta_{ij}-\frac{x_j}{N})x_i
\end{equation}

We search for a solution of this equation satisfying detailed balance
(i.e. $P^{st}f_i=D\sum_j\partial_j(g_{ij}P^{st})$).
Defining the marginalized probability distribution function:

\begin{equation}
p(x)=\int_0^\infty \prod_{j \neq i} dx_j P^{st}(\underline{x})
\end{equation}

\noindent we can easily obtain an equation for $p(x)$.

\begin{equation}
 b\left(1-\frac{sx}{N}\right) p(x) = D \frac{d}{dx}\left[
 \left(x-\frac{x^2}{N}\right) p(x)\right]
\end{equation}

This equation can be easily solved, giving:

\begin{equation}
p(x) \propto
x^{\beta-1}\left(1-\frac{x}{N}\right)^{\beta(s-\frac{1}{N})-1} \qquad \beta=\frac{b}{D}
\end{equation}

Notice that this distribution correctly shows the monodominance behavior 
 $\delta(0)$ or  $\delta(N)$ in the limit $\beta \rightarrow 0$.
Finally, if we fix $\mu=\frac{\beta s}{N}$, in the limit for $N \rightarrow
\infty$  we obtain:

\begin{equation}\label{soluzione}
p(x)=\frac{\mu^\beta}{\Gamma(\beta)x^{1-\beta}}e^{-\mu x} 
\end{equation}

In fig. 1 we plot simulation of the stationary p.d.f. for various value
of the parameter $\beta$,  and check the validity of (\ref{soluzione}).

\begin{figure}[htbp]
\includegraphics[width=8.5cm]{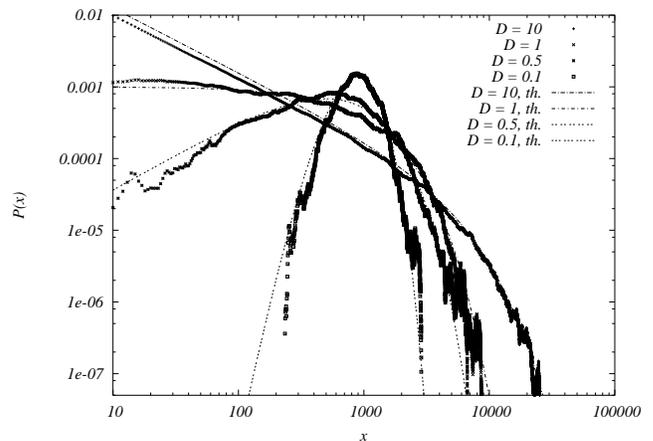}
\caption{simulation of marginalized stationary p.d.f. for various values of the
  diffusion coefficient $D$, compared with theoretical curves. For all
  curves $b=1$, $\bar{\lambda}=1$ $s=100$,$N=10^9$ Curves are binned
  linearly with binning size $\delta x =10^4$. Notice that as $D$
  increases the curve approaches the Fisher log series.
}
\end{figure}

Instead of having a system of stochastic differential equation, it is
possible to take into account the interaction of a species with the
ecosystem in an averaged way. Let us consider the Langevin equation:

\begin{equation}
\dot{x}(t) = b+ \bar{\lambda}x - \gamma x + D \sqrt{x} \xi
\end{equation}

\noindent where the parameter $\gamma$ takes into account the effect of
competition. In order to have normalizable solutions, we have to require
that $\gamma>\bar{\lambda}$. When this condition holds, it is straightforward to show
that the stationary p.d.f. satisfying detailed balance is the same as
(\ref{soluzione}), with $\mu=-(\bar{\lambda}-\gamma)/D$. Notice that in
this case, the detailed balance solution is exact; it is also remarkable
that the stationary distribution (\ref{soluzione}) can be achieved without fixing
neither the number of species, nor the number of individuals.

\section{Comparison with experimental data}

Among the most reliable data on single-trophic species distribution of species
abundance are tropical forest census \cite{condit}. In order to make a
coarse graining, a Preston plot is used: data are collected via a
logarithmic binning in base 2, and species at the edge between two consecutive
binning are equally divided between them. Since we have a continuous
probability density, we compared the histogram with the integral over the bins
of the distribution with the experimental data, and made a least-square
fit of the parameters $\beta$ and $\mu$, plus the normalization. We
found a good agreement of our predicted curve with the histogram; in
FIG.2 it is shown the comparison between our solution and the
lognormal. Notice that the
two distributions have the same number of fitted parameter. 
It would be interesting to compare
our distribution with data collected form other kind of ecosystems, and
to try to clarify the dependence of our free parameter $\beta$ from
ecological quantities like the immigration pressure, the speciation rate
and the extinction threshold.

\begin{figure}
\includegraphics[angle=-90,width=7.5cm]{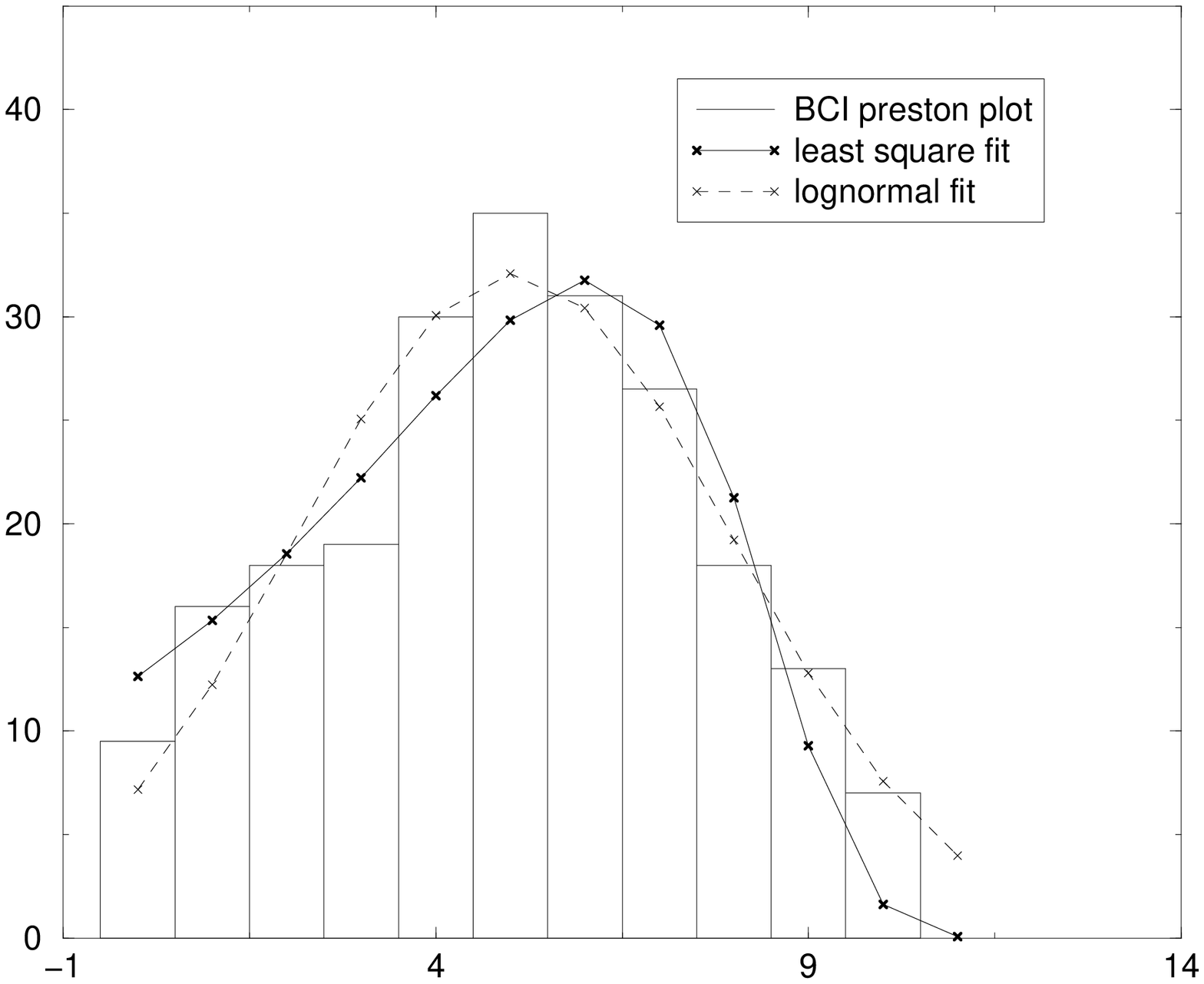}
\includegraphics[angle=-90,width=7.5cm]{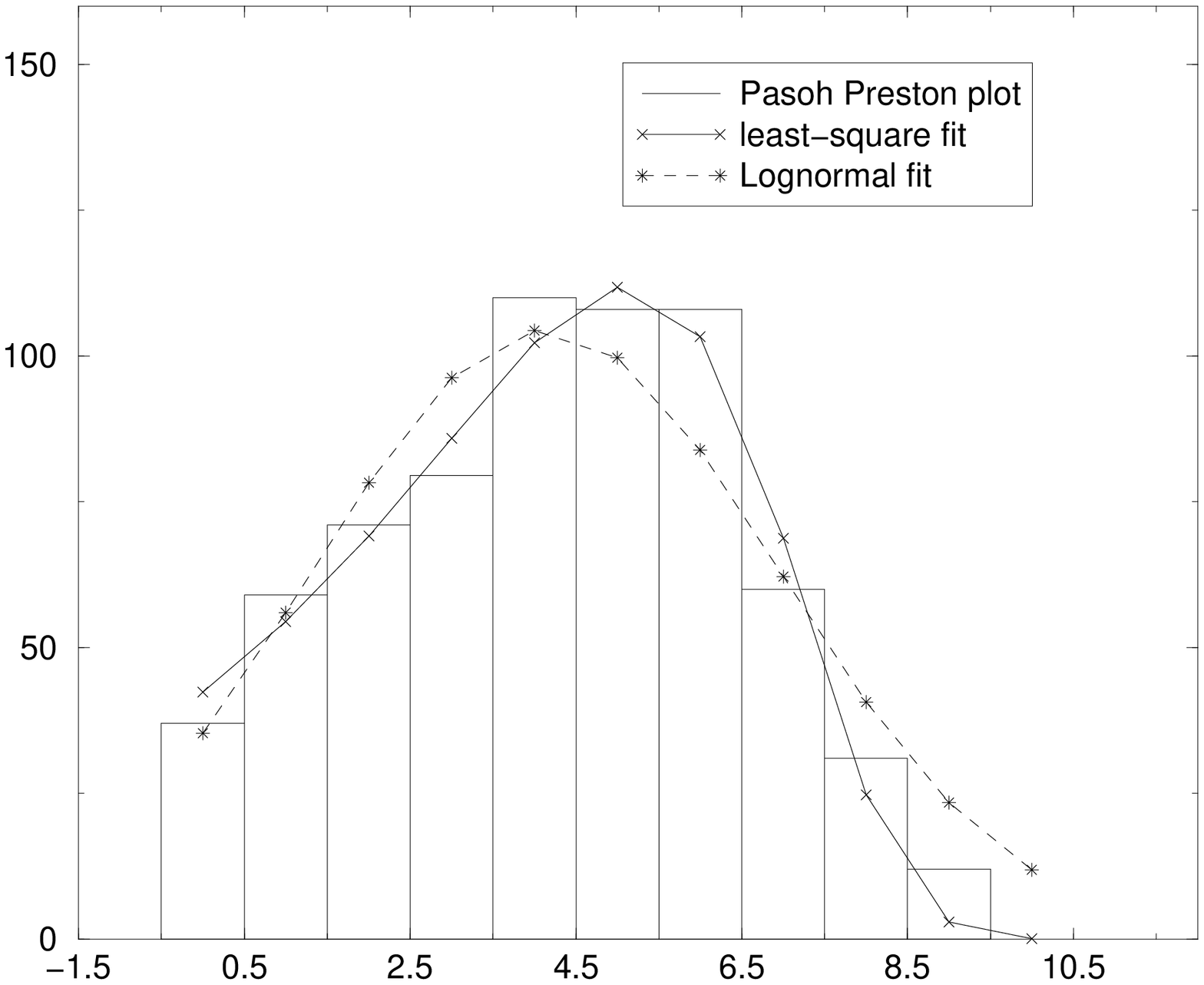}
\caption{fit of Barro Colorado Island and Pasoh specie abundance data - Preston
  plot \cite{hubbel}. Comparison between our solution and
  lognormal. Fitted value of the parameters of our distribution are
  $\beta = 0.23$ and $\mu = 0.010$ for the BCI; $\beta = .37$ and $\mu = 0.015$ for
  Pasoh. In absence of an objective estimate of the error bars on the
  observational data, both our result and  the lognormal give a
  reasonable fit.}

\end{figure}

\section{Discussion and Perspectives}

The model we introduce admits a family of stationary p.d.f. depending on
the parameter $\beta$.
This parameter fully determines the shape of the distribution: for
$\beta << 1$ one recovers the Fisher log series, while for $\beta$
large, one obtains a lognormal-like distribution. As we already pointed
out, both these distributions are well known in the population biology
literature as possible candidate to be the `right' distributions found in nature.

There is some analogy between our model and the Kesten process.
Indeed, also the Kesten process admits two different
regimes, one lognormal and one with a power law tail.
The main differences is that in our case the multiplicative random
process is applied to the square root of the variables, rather than to
the variable itself. As a consequence, in the Kesten case, the
exponent of the power law tail of the stationary distribution is always
greater than one, while  the small $\beta$ regime of our system is
characterized by a  power law tail over many decades, with an exponent
that is always less than $1$: the cutoff due to the conserved number of
individuals ensures the normalization of these long-tailed distributions. 

It is remarkable that our distribution is the same found in studies made
by Kerner in the '50 \cite{kerner} on the invariant
measure in a system of Lotka-Volterra equations with purely
asymmetric couplings. In that works the interactions are only of
predator-prey type, and the system is deterministic, while we are
considering a stochastic system with purely competitive coupling. The
discover of the same distribution in such different models suggests that
it might exist some deeper and more general mechanism determining the
statistical behavior of ecosystems, regardless of the type of 
interactions among species.

\begin{acknowledgments}
This work is a byproduct of many discussions with
J.Banavar and I.Volkov. 
\end{acknowledgments}

\end{document}